\documentstyle[12pt]{article}
\newcommand{\dI}{{\rm d}I}
\newcommand{\dx}{{\rm d}x}
\newcommand{\dy}{{\rm d}y}

\newfont{\Mb}{msbm10}

\begin{document}
\setcounter{equation}{0}
\setcounter{figure}{0}
\setcounter{table}{0}

\title{Solving second order ordinary differential equations by extending
       the PS method}
\author{
L.G.S. Duarte\thanks{
     Universidade do Estado do Rio de Janeiro,
     Instituto de F\'{\i}sica, Departamento de  F\'{\i}sica Te\'orica,
     R. S\~ao Francisco Xavier, 524, Maracan\~a, CEP 20550--013,
     Rio de Janeiro, RJ, Brazil. E-mail: lduarte@dft.if.uerj.br},
L.A. da Mota\thanks{
     idem. E-mail: damota@dft.if.uerj.br}
and J.E.F.Skea\thanks{
     idem. Email: jimsk@dft.if.uerj.br}
}

\maketitle

\abstract{An extension of the ideas of the Prelle-Singer procedure
to second order differential equations is proposed. As in the
original PS procedure, this version of our method deals with differential
equations of the form $y''={M(x,y,y')}/{N(x,y,y')}$, where $M$ and $N$
are polynomials with coefficients in the field of complex numbers $\it C$.
The key to our approach is to focus not on the final solution but on the
first-order invariants of the equation. Our method is an attempt to
address algorithmically the solution of SOODEs whose first integrals
are elementary functions of $x$, $y$ and~$y'$.
}

\newpage
\section{Introduction}

The fundamental position of differential equations (DEs) in scientific
progress has, over the last three centuries, led to a vigorous search
for methods to solve them. The overwhelming majority of these methods
are based on classification of the DE into types for which a method
of solution is known, which has resulted in a gamut of methods that
deal with specific classes of DEs. This scene changed somewhat at the
end of the 19th century when Sophus Lie developed a general method to
solve (or at least reduce the order of) ordinary differential equations
(ODEs) given their symmetry transformations ~\cite{step,bluman,olver}.  Lie's method
is very powerful and highly general, but first requires that we find
the symmetries of the differential equation, which may not be easy to do.
Search methods have been developed~\cite{nosso,nosso2} to extract
the symmetries of a given ODE, however these methods are heuristic and
cannot guarantee that, if symmetries exist, they will be found.

On the other hand in 1983 Prelle and Singer (PS) presented a deductive
method for solving first order ODEs (FOODE) that presents a solution
in terms of elementary functions if such a solution exists~\cite{PS}.
The attractiveness of the PS method lies not only in its basis on a
totally different theoretical point of view but, also in the fact that,
if the given FOODE has a solution in terms of elementary functions, the
method guarantees that this solution will be found (though, in principle
it can admittedly take an infinite amount of time to do so). The original
PS method was built around a system of two autonomous FOODEs of the form
$\dot{x} = P(x,y)$, $\dot{y}=Q(x,y)$  with $P$ and $Q$ in ${\it C}[x,y]$ or,
equivalently, the form $y'=R(x,y)$, with $R(x,y)$ a rational function
of its arguments.  Here we propose a generalization that allows us to
apply the techniques developed by Prelle and Singer to second order
differential equations (SOODEs).  The key idea is to focus not on the
final solution of the equation, but rather its invariants.

This paper is organized as follows: in section~\ref{PSreview}, the
reader is introduced to the PS procedure; section~\ref{ourmethod}
addresses our approach extending the ideas of the PS procedure to the
case of SOODEs and discusses how generally applicable the method is
to such equations.  Section~\ref{examples}
is dedicated to some examples solved via our procedure and, finally,
conclusions are presented in section~\ref{conclusion}.

\section{The Prelle-Singer Procedure}
\label{PSreview}

Despite its usefulness in solving FOODEs, the Prelle-Singer procedure is
not very well known outside mathematical circles, and so we present
a brief overview of the main ideas of the procedure.

Consider the class of FOODEs which can be written as
\begin{equation}
y' = {\frac{dy}{dx}} = {\frac{M(x,y)}{N(x,y)}}
\label{FOODE}
\end{equation}
where $M(x,y)$ and $N(x,y)$ are polynomials with coefficients in the complex 
field $\it C$.

In~\cite{PS}, Prelle and Singer proved that, if an elementary first
integral of~(\ref{FOODE}) exists, it is possible to find an integrating
factor $R$ with $R^n~\in~\it C$ for some (possible non-integer) $n$, such that
\begin{equation}
{\frac{\partial (RN)}{\partial x}}+{\frac{\partial (RM)}{\partial y}} = 0.
\label{eq_int_factor}
\end{equation}

The ODE can then be solved by quadrature.  From~(\ref{eq_int_factor})
we see that
\begin{equation}
  N\, {\frac{\partial R}{\partial x}}
+ R\, {\frac{\partial N}{\partial x}}
+ M\, {\frac{\partial R}{\partial y}}
+ R\, {\frac{\partial M}{\partial y}} 
= 0.
\label{eq_int_factor_aberta}
\end{equation}

Thus
\begin{equation}
{\frac{D[R]}{R}} =  - \left( {\frac{\partial N}{\partial x}} + 
{\frac{\partial M}{\partial y}} \right),
\label{eq_PS}
\end{equation}
where
\begin{equation}
\label{eq_def_D}
D \equiv N {\frac{\partial }{\partial x}}
        + M {\frac{\partial }{\partial y}}.
\end{equation}

Now let $R = \prod_i f^{n_i}_i$ where $f_i$ are irreducible polynomials 
and $n_i$ are non-zero integers. From (\ref{eq_def_D}), we have
\begin{eqnarray}
\label{ratio}
{\frac{D[R]}{R}} & = & {\frac{D[\prod_{i} f^{n_i}_i]}{\prod_i f^{n_k}_k}} =
                    {\frac{\sum_i f^{n_i-1}_i n_i D[f_i] \prod_{j \ne i}
                     f_j^{n_j}}{\prod_k f^{n_k}_k}} \nonumber \\[3mm]   
                 & = &  \sum_i {\frac{f^{n_i-1}_i n_i D[f_i]}{f_i^{n_i}}} =
                    \sum_i {\frac{n_iD[f_i]}{f_i}}.
\end{eqnarray}

From~(\ref{eq_PS}), plus the fact that $M$ and $N$ are polynomials, 
we conclude that ${D[R]}/{R}$ is a polynomial. Therefore,
from~(\ref{ratio}), we see that $f_i | D[f_i]$.

We now have a criterion for choosing the possible $f_i$ (build all
the possible divisors of $D[f_i]$) and, by using~(\ref{eq_PS})
and~(\ref{ratio}), we have
\begin{equation}
\label{eq_ni}
\sum_i {\frac{n_iD[f_i]}{f_i}} = -\left( {\frac{\partial N}{\partial 
x}} + {\frac{\partial M}{\partial y}} \right).
\end{equation}

If we manage to solve~(\ref{eq_ni}) and thereby find $n_i$,
we know the integrating factor for the FOODE and the problem is
reduced to a quadrature. Risch's algorithm~\cite{Risch} can then
be applied to this quadrature to determine whether a solution
exists in terms of elementary functions.

\section{Extending the Prelle-Singer Procedure}
\label{ourmethod}

In the previous section, the main ideas and concepts used in the
Prelle-Singer procedure were introduced. Here we present an extension of 
these ideas applicable to SOODEs. The main idea is to focus on
the first order invariants of the ODE rather than on the solutions.

\subsection{Introduction}
Consider the SOODE
\begin{equation}
\label{2ode}
y'' = {\frac{d^2y}{dx^2}} = {\frac{M(x,y,y')}{N(x,y,y')}},
\end{equation}
where $M(x,y,y')$ and $N(x,y,y')$ are polynomials with coefficients in~$\it C$.
We assume that~(\ref{2ode}) has a solution in terms of elementary
functions, in which case there are two independent elementary
functions of $x$ $y$ and $y'$ which are constant on all solutions
of~(\ref{2ode}), namely the first order invariants
\begin{equation}
I_i(x,y,y') = C_i \quad\quad i=1,2.
\end{equation}

\noindent
Without loss of generalization we consider one of these and, dropping
the index on $I_i$ we have
\begin{equation}
\label{diffI}
\dI = {\frac{\partial I}{\partial x}}\,\dx
     +{\frac{\partial I}{\partial y}}\,\dy
     +{\frac{\partial I}{\partial y'}}\,\dy' = 0.
\end{equation}
Now, introducing the notation ${\frac{\partial I}{\partial u}} \equiv I_{u}$,
we have
\begin{equation}
I_x+I_y y'+ I_{y'}y'' = 0,
\end{equation}
\noindent
and so
\begin{equation}
y'' = - {\frac{I_x+I_y y'}{I_{y'}}},
\end{equation}
which is~(\ref{2ode}) in terms of the differential invariant $I$.
Rewriting~(\ref{2ode}) as
\begin{equation}
\label{1form}
{\displaystyle \frac{M}{N}}\, \dx -  \dy' = 0
\end{equation}
and observing that
\begin{equation}
y' \,\dx = \dy, 
\end{equation}
we can add the identically null term $S(x,y,y') y'\,\dx - S(x,y,y')\,\dy$
to~(\ref{1form}) and obtain the 1-form
\begin{equation}
\label{1form2}
\left({\displaystyle \frac{M}{N}} +  S y'\right)\,\dx - S\,\dy -\dy' = 0.
\end{equation}

Notice that the 1-form~(\ref{1form2}) must be proportional to the 1-form
(\ref{diffI}). So, since the 1-form~(\ref{diffI}) is exact, we can 
multiply~(\ref{1form2}) by the integrating factor $R(x,y,y')$ to 
obtain
\begin{equation}
\label{1form2}
\dI = R (\phi +  S y')\,\dx - R S\,\dy - R\,\dy' = 0,
\end{equation}
where $\phi\equiv M/N$.

Comparing equations~(\ref{diffI}) and~(\ref{1form2}),
\begin{eqnarray}
\label{eqI}
I_x & = & R (\phi +  S y'), \nonumber \\
I_y & = & - R S, \nonumber \\
I_{y'} & = & - R.
\end{eqnarray}
Now equations~(\ref{eqI}) must satisfy
the compatibility conditions $I_{xy} = I_{yx}, I_{xy'} = I_{y'x}$
and $I_{yy'} = I_{y'y}$. This implies that
\begin{eqnarray}
\label{eqSA1}
D[S]  & = & - \phi_y + S \phi_{y'} + S^2,  \\
\label{eqSA2}
D[R]  & = & -R (S + \phi_{y'}),  \\
\label{eqSA3}
R_y & = & R_{y'} S + S_{y'} R,
\end{eqnarray}
where the differential operator $D$ is defined as
\begin{equation}
\label{eq_def_D2}
D \equiv {\frac{\partial }{\partial x}}
    + y' {\frac{\partial }{\partial y}}
   + \phi {\frac{\partial }{\partial y'}}.
\end{equation}

Combining~(\ref{eqSA1}) and~(\ref{eqSA2}) we obtain
\begin{equation}
D[RS]  =  -R \phi_{y}.
\label{DRS}
\end{equation}

So if the product of $S$ and the integrating factor $R$ is a rational 
function of $x$, $y$ and $y'$, then $D[RS]$ is too. Since $\phi$ is rational
(and so, therefore, is $\phi_{y}$), equation~(\ref{DRS}) tells us that $R$
is rational.  Using~(\ref{eqSA2}) and similar arguments we conclude that
$S$ must be a rational function of $x$, $y$ and $y'$.

In summary, from~(\ref{eqI}) it follows that the supposition that $RS$ is
rational can be equated to the existence of a first order invariant
whose derivatives in relation to $x$, $y$ and $y'$ are rational functions.
With  this in mind we restate the original supposition in the form of a
conjecture.

\subsection{The Conjecture}
\label{conjecture}

We first state a result proved in~\cite{PS}:
\bigskip
\noindent
{\bf Theorem:}{\em
Let $K$ be a differential field of functions in $n+1$ variables and $L$
an elementary extension of $K$. Let $f$ be in $K$ and assume there exists
a nonconstant $g$ in $L$ such that $g$ is constant on all solutions of
$y^{(n)} = f(x,y,y',y'',\ldots,y^{(n-1)})$. Then there exist
$w_1,\ldots,w_m$ algebraic over $K$ and constants $c_i,\ldots,c_m$ such that
\begin{equation}
\label{19}
w_0(x,y,y',y'',..,y^{(n-1)})
+ \sum_i c_i \log(w_i(x,y,y',y'',\ldots,y^{(n-1)}))
\end{equation}
is a constant on all solutions of $y^{(n)} = f(x,y,y',y'',\ldots,y^{(n-1)})$.
}

\bigskip
This result shows that for the particular case of SOODEs whose solutions 
are elementary, there are two independent first order invariants of the 
form
\begin{equation}
\label{diffinv}
w_0(x,y,y') + \sum_i c_i \log w_i(x,y,y').
\end{equation}

Our conjecture is that if these two first order invariants exist it is
always possible to find a function of them
(which will, therefore, itself be a first order invariant) of the form

\begin{equation}
z_0(x,y,y') + \sum_i c_i \log[z_i(x,y,y')],
\end{equation}
where $z_i$ are {\em rational\/} functions of $x$, $y$ and $y'$.

\bigskip
\noindent
{\bf Conjecture:}{\em
Let $K$ be a differential field of functions in three variables and $L$ an 
elementary extension of $K$. Let $f$ be in $K$ and assume there exist two
independent nonconstant $\{g_1,g_2\}$ in $L$ such that $g_i$ are constant
on all solutions of $y'' = f(x,y,y')$.
Then there exists at least one constant of the form
\begin{equation}
\label{eq_conjecture}
z_0(x,y,y') + \sum_i c_i \log(z_i(x,y,y'))
\end{equation}
where the $z_i$ are in $K$.}

\bigskip
By the previous reasoning it can be seen that~(\ref{eq_conjecture}) implies
that the product $RS$ is a rational function of $x$, $y$ and $y'$.

If this conjecture holds, then our extension of the PS method applies to
all SOODEs of the form~(\ref{2ode}). Though we have not been able to
prove our conjecture, extensive trials while developing this procedure
has not revealed any counter example. Even if the conjecture is false,
our experience with real test cases has shown that the method is, at
least, applicable to the vast majority of SOODEs of the
form~(\ref{2ode}).

\subsection{Finding $R$ and $S$}

Our conjecture implies that, if the SOODE to be solved has an 
elementary general solution, then $S$ is a rational function which
we may write as
\begin{equation}
S = {\frac{S_n}{S_d}} = {\frac{\sum_{i,j,k} a_{ijk}{x}^{i} {y}^{j} 
{y'}^{k}}
{\sum_{i,j,k} b_{ijk}{x}^{i} {y}^{j} {y'}^{k}}}.
\end{equation}

We can also see that~(\ref{eqSA1}) does not involve $R$. So, given
a degree bound on the polynomials $S_n$ and $S_d$, we may find a set
of solutions to this
equation which are then candidates to solve the system of equations
(\ref{eqSA1})--(\ref{eqSA3}).

From~(\ref{eqSA2}) we have
\begin{equation}
{\frac{D[R]}{R}} = - (S + \phi_{y'}) = - {\frac{S_n}{S_d}}-
\left({\frac{M}{N}}\right)_{y'} = - {\frac{S_n N^2+S_d (N M_{y'}-M 
N_{y'})}{S_d N^2}}
\end{equation}
which can be rewritten as
\begin{equation}
\label{eq_PSext}
{\frac{{\cal D}[R]}{R}} = - S_n N^2+S_d (N M_{y'}-M N_{y'}),
\end{equation}
where the differential operator ${\cal D}$ is defined as
\begin{equation}
{\cal D} \equiv (S_d N^2) D.
\end{equation}

We keep in mind that
\begin{itemize}
\item $S_n$, $S_d$, $N$ and $M$ are polynomials in $x$, $y$ and $y'$;
\item ${\cal D}$ is a linear differential operator whose coefficients 
of
${\frac{\partial}{\partial x}}$, ${\frac{\partial}{\partial y}}$
and ${\frac{\partial}{\partial y'}}$ are polynomials in $x$, $y$ and $,y'$;
\item $R$ is a rational function of $x$, $y$ and $y'$, which we may write
as
\begin{equation}
R = {\frac{R_n}{R_d}} = {\frac{\sum_{i,j,k} c_{ijk}{x}^{i} {y}^{j} 
{y'}^{k}}
{\sum_{i,j,k} d_{ijk}{x}^{i} {y}^{j} {y'}^{k}}}.
\end{equation}
\end{itemize}

If we have a theoretical limit on the degrees of $R_m$ and $R_d$
(a {\em degree bound\/}), we may use a procedure 
analogous to that described in section~\ref{PSreview}
to obtain candidates for the integrating factor $R$. We simply
construct all polynomials in $x$, $y$ and $y'$ up to the degree bound.

\subsection{Reduction of the SOODE}

Once $R$ and $S$ have been determined using equations~(\ref{eqI}) we 
have all the partial first derivatives of the first order differential
invariant, $I(x,y,y')$,  which is constant on the solutions.
This invariant can then be obtained as

$$I(x,y,y') = \int \!R\left (\phi+S{\it y'}\right ){dx} \,\,- $$
$$\int \! \hbox{\large [} RS+{\frac {\partial}{\partial y}}\int
\!R\left (\phi+S{\it y'}\right ){dx} \hbox{\large ]} {dy} \,\,- $$
\begin{equation}
\label{C1}
\int \!\left[R+
{\frac {\partial }{\partial {\it y'}}}\left (\int \!R\left (\phi+S{
\it y'}\right ){dx}-\int \!\hbox{\large [}RS+{\frac {\partial }{\partial 
y}}\int \!R
\left (\phi+S{\it y'}\right ){dx}\hbox{\large ]}{dy}\right 
)\right]{d{\it y'}}.
\end{equation}
\noindent
The equation $I(x,y,y') = C_1$ can then be solved to obtain a FOODE
for $y'$: the {\em reduced\/} ODE
\begin{equation}
\label{reduced_ode}
y' = \varphi (x,y,C1).
\end{equation}

\noindent 
To obtain the general solution of the original ODE, we can apply the
Prelle-Singer method in its original form to this reduced ODE. Thus, if our
conjecture is correct, the method proposed here (for SOODEs of the
form~(\ref{2ode})) is as algorithmic as the original PS method for FOODEs. We
note that the original PS method fails to be what is strictly an algorithm
because no theoretical degree bound is yet known for the candidate polynomials
which enter in the prospective solution, and so the procedure has no effective
terminating condition for the case when an elementary solution does not exists.
In practice, a terminating condition is put in by hand (it is found that
polynomials of degree higher than 4 lead to computations which are overly
complex for the average desktop computer). However, should such a degree bound
be established, and our conjecture shown to be true, then the method proposed
here would be an algorithm for deciding whether elementary solutions of SOODEs
of the form~(\ref{2ode}) exist.

\section{Examples}
\label{examples}

In this section we present examples of physically motivated SOODEs that are
solved by our procedure\footnote{We present only the reduction of the SOODEs
since the integration of the resulting FOODE can be achieved by various
methods, including the PS method itself.}. As a simple illustrative example,
we begin with the classical harmonic oscillator and then consider some
nonlinear SOODEs which arise from astrophysics and general relativity.

{\bf Example 1: The Simple Harmonic Oscillator}\\
In its simplest form, the equation for the simple harmonic oscillator is
\begin{equation}
\label{HO}
y'' = -y.
\end{equation}

For this ODE equations (\ref{eqSA1}), (\ref{eqSA2}) and (\ref{eqSA3}) are
\begin{eqnarray}
S_x + y' S_y - y S_{y'} & = & 1 + S^2,  \\
R_x + y' R_y - y R_{y'}  & = & -R S,   \\
R_y - R_{y'} S - S_{y'} R& = & 0.
\end{eqnarray}
One possible solution to these equations is
\begin{equation}
S = {\frac{y}{y'}}, \,\,\, R = y'.
\end{equation}

From this, and using~(\ref{C1}), we get the reduced ODE
\begin{equation}
C1 = y^2 + y'^{2},
\end{equation}
which, of course,  represents the energy conservation for the oscillator.

This example is very simple and leads to a form of $\phi$ which is
independent of $x$ and $y'$. And, as with all linear ODEs, alternative
and more straightforward solution methods exist. The other examples
illustrate the solution method at work for non-linear SOODEs which
can be placed in the form~(\ref{2ode}).

{\bf Example 2: An Exact Solution in General Relativity}

A rich source of non-linear DEs in physics are the highly non-linear
equations of General Relativity. In general, Einstein's equations are,
of course, partial DEs, but there exist classes of equations where the
symmetry imposed reduces these equations to ODEs in one independent
variable. One such class is that of static, spherically symmetric
solutions for stellar models, which depend only on the radial variable,
$r$. The metric for a general statically spherically spacetime has
two free functions, $\lambda(r)$ and $\mu(r)$ say. On imposing the condition
that the fluid is a perfect fluid, Einstein's equations reduce to two coupled
ODEs for $\lambda(r)$ and $\mu(r)$. Specifying one of these functions
reduces the problem to solving an ODE (of first or second order) for the
other.

Following this procedure, Buchdahl~\cite{Buchdahl1} obtained an exact
solution for a relativistic fluid sphere by considering the so-called
isotropic metric
\[
\dot{s}^2 =
  (1-f)^2(1+f)^{-2}\dot{t}^2
-(1+f)^4[\dot{r}^2+r^2(\dot{\theta}^2+\sin^2\theta\,\dot{\phi}^2]
\]
with $f=f(r)$. The field equations for $f(r)$ reduce to
\[
  ff''-3f'^2-r^{-1}ff' = 0.
\]
Changing notation with $y(x)=f(r)$, equations (\ref{eqSA1}), (\ref{eqSA2})
and (\ref{eqSA3}) assume the form
\begin{eqnarray}
S_x + y' S_y + {\frac {{\it y'}\,\left (3\,{\it y'}\,x+y\right )}{xy}} 
S_{y'} & = & - {\frac {{\it y'}}{xy}} + {\frac {{\it y'}\,\left 
(3\,{\it y'}\,x+y
\right )}{x{y}^{2}}} + \nonumber \\
&& \left( {\frac {3\,{\it y'}\,x+y}{xy}}+3\,{\frac {{\it y'}}{y}} \right) S + S^2,  \\
R_x + y' R_y + {\frac {{\it y'}\,\left (3\,{\it y'}\,x+y\right )}{xy}} R_{y'}  & = &-R\left (S+
{\frac {3\,{\it y'}\,x+y}{xy}}+3\,{\frac {{\it y'}}{y}}
\right )
,  \\
R_y - R_{y'} S - S_{y'} R& = & 0.
\end{eqnarray}

One solution of those equations is
\begin{equation}
S = -3\,{y'\over y},~~~ R = {1\over xy^3}
\end{equation}
By using~(\ref{C1}) we obtain the reduced FOODE:
\begin{equation}
C_1 =y'/(y^3x)
\end{equation}
which is separable and easily integrated to obtain the general solution

\begin{equation}
y(x)^{2}=\left (-{ C_1}\,{x}^{2}+{\it C_2}\right )^{
-1}.
\end{equation}

{\bf Example 3: A Static Gaseous General-Relativistic Fluid Sphere}

In a later paper~\cite{Buchdahl2}, Buchdahl approaches the problem
of the general relativistic fluid sphere using a different coordinate
system from the previous example. For ease in comparison of the originals,

Substituting the $\xi(r)$ of the original
Writing $y(x)$ instead of the $\xi(r)$ in the original,
arrives at the equation
\begin{equation}
y'' = {\frac {{x}^{2}{y'}^{2}+{y}^{2}-1}{{x}^{2}y}},
\end{equation}

For this SOODE, eqs (\ref{eqSA1}, \ref{eqSA2} and \ref{eqSA3}) become:

\begin{eqnarray}
S_x + y' S_y + {\frac {{x}^{2}{y'}^{2}+{y}^{2}-1}{{x}^{2}y}} 
S_{y'} &
= & -2\,{x}^{-2}+{\frac {{x}^{2}{y'}^{2}+{y}^{2}-1}{{y}^{2}{x}^{2}}} 
\nonumber \\
&&+2\,{\frac {y'\,S}{y}}+{S}^{2}  \\
R_x + y' R_y + {\frac {{x}^{2}{y'}^{2}+{y}^{2}-1}{{x}^{2}y}} R_{y'}  & = & -R\left (S+2\,{\frac 
{y'}{y}}\right )  \\
R_y - R_{y'} S - S_{y'} R& = & 0
\end{eqnarray}

One solution to those equations is:

\begin{equation}
S = {\frac {-{x}^{2}{y'}^{2}-xyy'+1}{x{y}^{2}+{x}^{2}yy'
}}
, \,\,\, R = {\frac {y+xy'}{x{y}^{2}}}.
\end{equation}

From this, using eq. (\ref{C1}), we get the reduced FOODE:

\begin{equation}
C_1 = {\frac {2\,xyy'+{y}^{2}+{x}^{2}{y'}^{2}-1}{2{x}^{2}{y }^{2}}}
.
\end{equation}

which can be solved to:
\begin{equation}
y(x)^{2}={\frac {\tan(\sqrt {2}\sqrt {{\it C_1}}
\left ({\it C_2}+x\right ))^{2}}{\left (2\,{\it C_1}+2\,
\tan(\sqrt {2}\sqrt {{\it C_1}}\left ({\it C_2}+x\right ))^{2
}{\it C_1}\right ){x}^{2}}}
\end{equation}

This example has an extra feature: It is not solved by  other solvers we 
have
tried (mainly the Maple solver, in the version 5, that we believe to be 
the
best). So, apart from the (already) very interesting fact that our 
approach is an
algorithmic attempt to solve SOODEs, we have also this present fact, 
i.e., some
SOODEs are solved via our method and ``escape'' from other very powerful
solvers.

\section{Conclusion}
\label{conclusion}
\

In this paper, we presented an approach that is an extension of the 
ideas
developed by Prelle-Singer \cite{PS} to tackle FOODEs. We believe it to be 
the first
technique to address algorithmically the solution  of SOODEs with 
elementary
first integrals.

Here, we dealt with a restrict class of SOODEs (namely, the ones of the
form (\ref{2ode})). However, we can use our method in solving SOODEs where
$\phi(x,y,y')$ depends on elementary functions of $x,y,y'$, following 
the
developments for the Prelle-Singer approach for FOODEs \cite{Man1,Man2}. We are 
presently
working on those ideas.

The generality of our approach is based on a conjecture (see section 
(\ref{conjecture}))
that we have already proved for many special cases. Even if the conjecture is proven 
false,
our approach is a powerful tool in dealing with SOODEs since we have 
extensively
tested it with many equations, both from mathematics and physical 
origin. In fact,
since all the examples we have encounter have been solved by our 
approach,
we are preparing a computational package implementing the Prelle-Singer
procedure (and our present extension) to be submitted to Computer 
Physics
Communications.


\end{document}